\documentclass[aps,floatfix,superscriptaddress,notitlepage]{revtex4}
\usepackage{amsmath,amssymb,amsfonts,graphics,graphicx,dcolumn,bm,enumerate}
\usepackage{comment,natbib,appendix}
\usepackage[ddmmyyyy,hhmmss]{datetime}
\usepackage{multirow,color}
\usepackage{amsthm}
\usepackage{epsfig}
\usepackage{natbib}
\usepackage{hyperref}
\usepackage{epstopdf}

\newcommand{\cmp}
{\affiliation{Condensed Matter Physics Division, 
Saha Institute of Nuclear Physics, 1/AF Bidhannagar, Kolkata 700064, India.}}
\newcommand{\isi}
{\affiliation{Economic Research Unit, Indian Statistical Institute, 203 B. T. 
Road, Kolkata 700108, India.}}

\begin{document}

% \begin{tiny}updated at \currenttime ~on \today                                  
% \end{tiny}

\title{Fat tailed distributions for deaths in conflicts and disasters}

\author{Arnab Chatterjee}%
\email[Email: ]{arnabchat@gmail.com} 
\cmp
\author{Bikas K Chakrabarti}%
\email[Email: ]{bikask.chakrabarti@saha.ac.in}
\cmp \isi

\begin{abstract}
We study the statistics of human deaths from wars, conflicts, similar man-made 
conflicts as well as natural disasters. 
The probability distribution of number of people killed in natural disasters as 
well as man made situations show power law decay for the largest sizes, with 
similar exponent values.
Comparisons with natural disasters, when event sizes are measured in terms of 
physical quantities (e.g., energy released in earthquake, volume of rainfall, 
land area affected in forest fires, etc.) also show striking resemblances.
The universal patterns in their statistics suggest that some subtle similarities 
in their mechanisms and dynamics might be responsible.
\end{abstract}

\maketitle

%%%%%%%%%%%%%%%%%%%%%%%%%%%%%%%%%%
\section{Introduction}
%%%%%%%%%%%%%%%%%%%%%%%%%%%%%%%%%%%%

Social dynamics is being studied extensively, qualitatively and 
quantitatively at 
present~\cite{Chakrabarti2006econosocio,stauffer2006biology,CFL_RMP2009,
galam2012sociophysics,Sen:2013}. 
This is an interdisciplinary research area, which has 
been a traditional ground for social scientists, but has  involved increasing 
participation of physicists in recent times, apart from mathematicians, 
computer scientists and others.
The behavior of social systems are quite interesting -- they 
demonstrate rich emergent phenomena, including `stunning 
regularities'~\cite{buchanan07}, which result out of interaction of a large 
number of entities (or agents)~\cite{liggett1999stochastic}. These rich 
dynamical systems can be studied well using tools of statistical 
physics~\cite{CFL_RMP2009,Sen:2013}.

The natural and physical sciences are rich with ``laws'' and universal 
patterns. There are not too many in socio-economic systems. The  Zipf's 
law~\cite{Zipf:1949} statistically models the frequency of words uttered in a 
given natural language, and is also observed to hold for city size, originally 
reported by Auerbach~\cite{auerbach1913} and well studied till 
date~\cite{gabaix1999zipf}. Zipf law has also been observed for size 
of firms, income distribution of 
companies, firm bankruptcy, etc.
Formally, the law states gives the frequency of the element ranked $k$ 
(according to size) as $f_k \sim 1/k^{\gamma}$. $\gamma$ ranges roughly between 
$0.8$ and $1.2$ for cities~\cite{gabaix2004evolution}, around $1$ for firm 
sizes~\cite{axtell2001zipf}, $0.7$ for income of companies~\cite{Okuyama1999} 
and $0.9$ for firm bankruptcy~\cite{Fujiwara2004}.
The corresponding probability density for the sizes of the elements is given by  
$P(s) \sim s^{-\nu} (\nu > 0)$ and the exponents are related as $\nu= 
1+\frac{1}{\gamma}$. Hence, the corresponding values of the size exponent $\nu$ 
are about $1.8-2.2$ for city size, $2$ for firm sizes, $2.4$ for company 
incomes and $2.1$ for firm bankruptcy. This is otherwise known as the Pareto 
law, as  and was first observed for the land wealth of the rich by 
Pareto~\cite{Pareto-book}, and subsequently known to hold for income and wealth 
distributions, with $\nu$ typically between $2$ and $3$ (see e.g., 
Ref.~\cite{chakrabarti2013econophysics}).
However, consumption expenditure distribution has been mostly reported to have 
a log-normal bulk with a power law 
tail~\cite{mizuno2008pareto,fagiolo2010distributional,chatterjee2016invariant,
chakrabarti2016quantifying}.
Social networks show certain empirical regularities: e.g., the number of links 
to websites have a power law 
tail~\cite{broder2000graph,albert2000error,adamic2000power}, 
degree distribution of movie actor network~\cite{barabasi1999emergence}
and sexual contacts~\cite{liljeros2001web} both show power law tails,
co-authorship networks~\cite{barabasi2002evolution} and 
citation distributions for articles have power law 
tails~\cite{Shockley:1957,Redner:2005} along with a lognormal 
bulk~\cite{Radicchi:2008,chatterjee2016universality}, among many others.

Several aspects of human lives have been well studied and documented, and 
certain regularities have already been established. For example, the most 
studied aspect is that of mortality rate.
The Gompertz-Makeham law of mortality describes the age dynamics of human 
mortality reasonably in the age window $30-80$ years, which states that the 
human death rate is the sum of an age-independent 
component~\cite{makeham1860law} and an age-dependent 
component~\cite{gompertz1825nature} which increases exponentially with age.

The history of human civilization has been shaped by events of disasters, wars 
and conflicts.
In recent times, the scale of disaster events have increased remarkably.
Growing population around the world has been seen as one of 
the reasons for the increase in counts of people affected by disaster events, 
as it has reported that more than $200$ million people have been affected by 
natural disasters every year since 1990~\cite{guha2011annual}, and a much 
larger 
number lives at risk of being affected.
Similarly, armed conflicts have at times threatened the public health of 
generations across the globe~\cite{leaning2013natural}. As civilization 
progressed, armed conflicts and wars have been instrumental in shaping the 
course of kingdoms and states. Our ancient epics (\textit{Mahabharata, 
Ramayana, Illiad} etc.) talk about wars of gigantic scale, and so do 
ancient historical documents and scriptures. Clashing kingdoms, clans and 
tribes, these wars lasted from days to years, resulting in fatality of enormous 
proportions. In the age of kingdoms, wars of different scales, small and large 
were common. In the recent global political scenario, conflicts happen at a 
rather smaller scale, except that the two \textit{World Wars} have seen large 
scale political polarization and military engagement resulting in fatalities 
of huge proportions.
Richardson's early 
works~\cite{richardson1948variation,richardson1960statistics} suggest that the 
fatality of wars decay as a power law: $P(x) \sim x^{-\alpha}$ for large values 
of fatality $x$. Subsequent careful analysis have confirmed these findings, 
e.g. the study of the Correlators of Wars (CoW) 
database~\cite{CoWdata} suggests an exponent $\alpha \simeq 
1.4$~\cite{cederman2003modeling}, and this suggests a heavy tailed distribution 
where the average is controlled by the deadliest of wars.
Wars have often been studied in the light of self-organized 
criticality~\cite{roberts1998fractality}.
Clauset et al. analyzed the probability distribution of terrorist attack deaths 
between 1968 and 2005 and found the decay exponent for the power law fit to be 
around $2.4-2.5$~\cite{clauset2007frequency}, while within shorter spans, the 
exponent had a large variation, $1.75-2.75$.

Disasters can be broadly classified into
 \textit{natural} and \textit{man-made}, while the natural disasters consist
 of the majority of recorded calamities. Natural disasters can be further 
classified into biological, climate-related or geophysical. The earth has seen 
three times as many natural disasters between 2000 and 2009 compared to that 
between 1980 and 1989~\cite{leaning2013natural}.
 What also probably plays a role here is the awareness in recording events,
 but however, what cannot be ignored is the growth in the climate related 
events, although the contribution of geophysical events have not changed much.
 The possible reasons for increase in events can be attributed to deforestation,
 increased urbanization and rapid industrialization, leading to
 increase in mean global temperature, massive precipitation and violent storms.
 A previous study on the total number of people affected and killed at various 
geographical locations have revealed that its probability distribution has a 
power law tail with exponent around $1.6-2.1$~\cite{becerra2012natural}.
Economists have also defined macroeconomic disasters, by studying the real 
personal consumer expenditure per capita or real GDP per capita, and found 
their size distribution to have a power law tail with very high exponents,
typically around $4$~\cite{barro2011size}.

In this paper, we analyze data for human deaths in \textit{man-made} conflicts 
like wars and battles, as well as from disasters. 
The data sources are given in the following section.
The probability distribution of the number of human deaths 
in a single event is observed to have a broad distribution with a power law 
tail. We argue that there might be similar mechanisms at play for the two 
phenomena that result in similar power law exponents.

\section{Results}
\label{sec:results}
\begin{figure}[t]
\includegraphics[width=17.5cm]{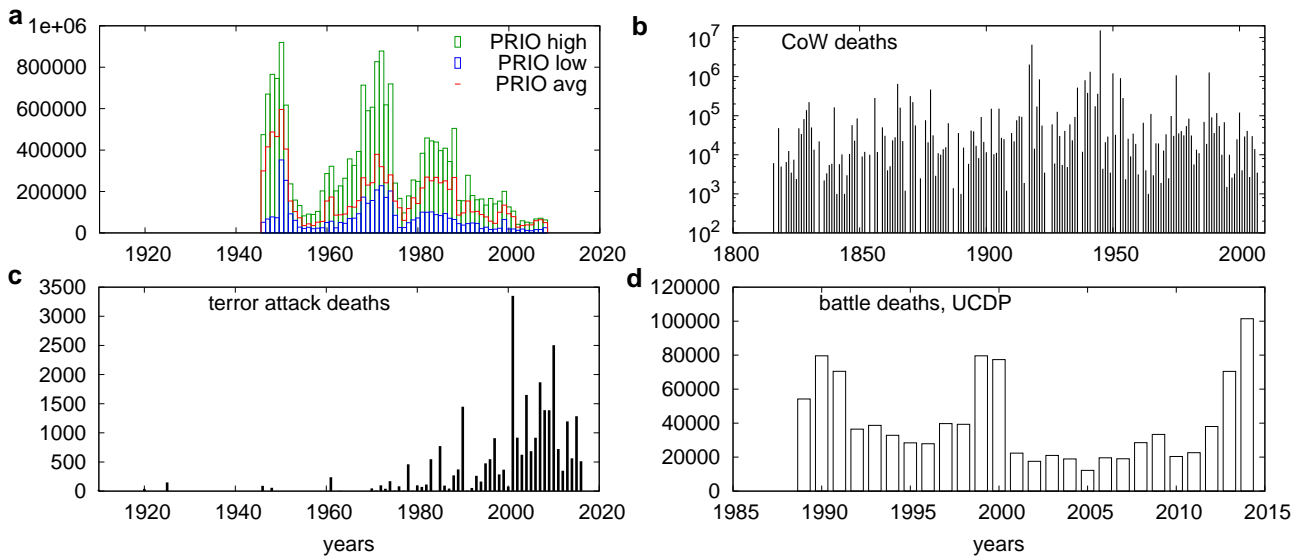}
 \caption{
 Statistics of fatalities over years for wars and conflicts: 
 (a) Number of people died in conflicts 
during 1946-2008, according to the PRIO database~\cite{PRIOdata}, with low, 
high and average estimates; 
(b) Number of human death according to the Correlators of Wars (CoW) 
database~\cite{CoWdata} during 1816-2007, consisting of Intra state, Inter 
state, Extra state and Non state wars;
(c) Number of human death in terror attack~\cite{Terrordatawiki} during 1910 
till July 
2016;
(d) Humber of human deaths in battles according to UCDP 
database~\cite{UCDPdata} during 1989-2014.
 }
 \label{fig:wardeath}
\end{figure}

\begin{figure}[t]
\includegraphics[width=17.5cm]{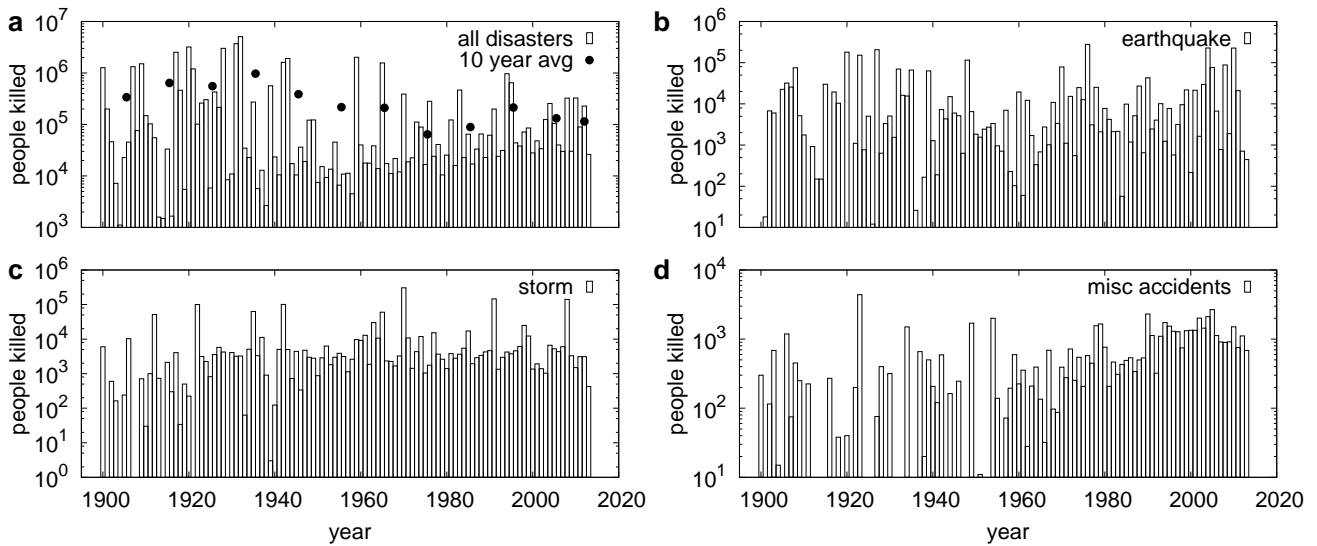}
 \caption{
 Statistics of fatalities over years for disasters during 1900-2013 according 
to the EMDAT database~\cite{EMDATdata}: 
 (a) Number of people died due to all disasters, along with 10 years averages 
indicating a decreasing trend with time. Number of human deaths due to (b) 
earthquake, (c) storms and (d) miscellaneous accidents in the same period.
 }
 \label{fig:disasterdeath}
\end{figure}

We access publicly available databases that keep count of the number of human 
deaths in wars and conflicts as well as disasters, over a large time span. 
We use the following datasets:
\begin{enumerate}
 \item  The Battle Deaths Dataset of Peace Research Institute Oslo 
(PRIO)~\cite{PRIOdata} for death in conflicts during 1946-2008.
Due to uncertainty in the exact number of fatalities, low, high and average 
estimates are provided for certain events. 
\item The war data~\cite{sarkees2010resort} from The Correlates of War Project 
(CoW)~\cite{CoWdata}
for deaths during 1816-2007. We use the number of deaths for Intra state, 
Inter state, Extra state and Non state wars.
\item  Battle-Related Deaths Dataset from Uppsala Conflict Data Program 
(UCDP)~\cite{UCDPdata} for deaths during 1989-2014,
\item Terror attacks data from List of battles and other violent events by 
death toll from 1910 till July 2016, Wikipedia~\cite{Terrordatawiki},
\item The international Disaster Database, EMDAT~\cite{EMDATdata} for 
fatalities from disasters during 1900-2013.
\end{enumerate}
The databases contained information about each single event of war, battle, 
conflict, terror attack, disaster, with the year of the event, as well as the 
number of people dead in each event. In some cases where the numbers are not 
clearly determined, low, high and possible average estimates are provided (e.g. 
PRIO data). In Fig.~\ref{fig:wardeath} we plot the number of deaths 
corresponding to different years from the different databases of wars, battles, 
conflicts and terror attacks. In Fig.~\ref{fig:disasterdeath} we plot the 
number of deaths corresponding to different years for all disasters, 
earthquakes, storms and miscellaneous accidents.

\begin{figure}[t]
\includegraphics[width=17.5cm]{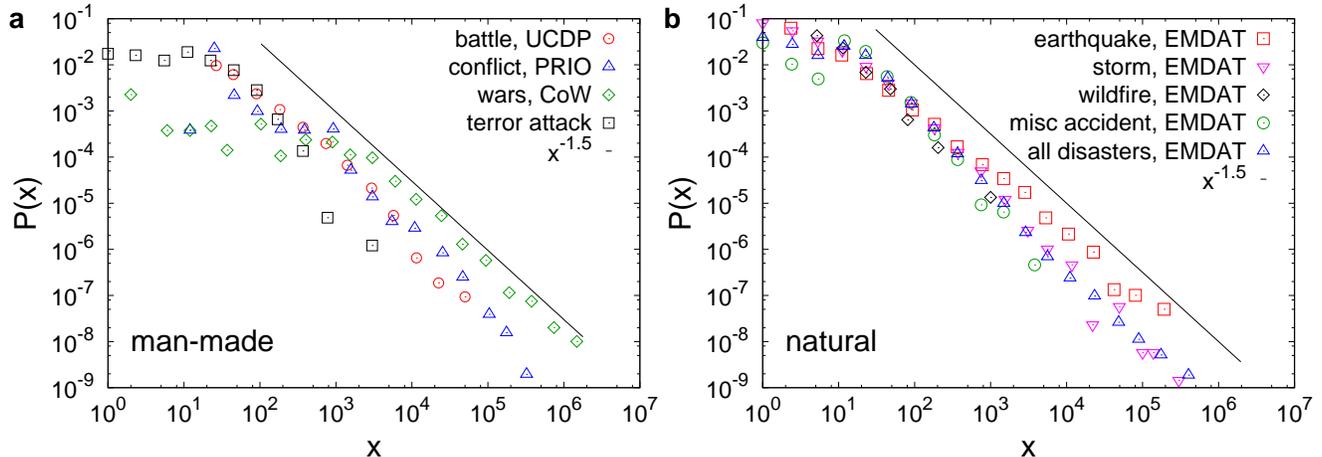}
 \caption{
The probability distributions $P(x)$ of event size $x$, measured by the number 
of human deaths in the corresponding event.
(a) \textit{Man-made events}: human deaths from conflicts 
during 1946-2008, according to the PRIO database~\cite{PRIOdata} (using lowest 
estimates),
dead according to the Correlators of Wars (CoW) 
database~\cite{CoWdata} during 1816-2007, 
dead in terror attack~\cite{Terrordatawiki} during 1910 till July 
2016, and battle deaths according to UCDP database~\cite{UCDPdata} during 
1989-2014.
Except the terrorist attack data, all these distributions seem to have a power 
law tail with similar exponents. The straight line is a guide to the exponent 
value $1.5$ for comparison. 
(b) \textit{Natural disasters}: human death from 
earthquakes, storms, wildfires, miscellaneous accidents, as well as all natural 
disasters listed in  the EMDAT database~\cite{EMDATdata} during 1900-2013.
The precise values of the estimated exponents are given in Table~\ref{tab:expo}.
 }
 \label{fig:alldeaths}
\end{figure}

We also plot the probability distributions $P(x)$ of 
events size $x$ measured by the number of deaths in the corresponding event.
In Fig.~\ref{fig:alldeaths}a, we plot the man made events, like wars, battles, 
conflicts and terror attacks. For each set, the probability distribution $P(x)$ 
seems to decay with a power law tail, with an exponents $1.54\pm0.06$ for PRIO, 
$1.63\pm0.03$ for CoW, and $1.64\pm0.07$ for UNDP.
In Fig.~\ref{fig:alldeaths}b, we plot the number of people killed in 
each event of natural disasters like earthquake, storms and miscellaneous 
accidents. The probability distribution $P(x)$ in these sets also seem to 
decay with a power law tail, with an exponent close to $1.51\pm0.05$, 
$1.65\pm0.03$, $1.51\pm0.01$ and $1.81\pm0.12$ respectively for earthquakes, 
storms, wildfires and miscellaneous accidents. The power law decay exponent for 
all disaster deaths (these also include mass movements, flood, drought, 
volcano, industrial reasons, epidemics, extreme temperatures, transport 
accidents, etc.) is around $1.48\pm0.03$.
We computed the power law exponents at the tail of the distributions, using the 
maximum likelihood estimates (MLE).
The precise exponent values are given in Table.~\ref{tab:expo}.

%%%%%%%%%%%%%%%%%%%%%%%%%%%%%%%%%%%%%%%%%%%%%%%%%%%%%%%%%%
%%%%%%%%%%%%%%%%%%%%%%%%%%%%%%%%%%%%%%%%%%%%%%%%%%%%%%%%%%%%
%%%%%%%%%%%%%%%%%%%%%%%%%%%%%%%%%%%%%%%%%%%%%%%%%%%%%%%%%%
%%%%%%%%%%%%%%%%%%%%%%%%%%%%%%%%%%%%%%%%%%%%%%%%%%%%%%%%%%%%

%%%%%%%%%%%%%%%%%%%%%%%%%%%%%%%%%%%%%%%%%%%%%%%%%%
\begin{table}[!htbp]
\caption{Dataset description and estimated value of power law exponent $\nu$ as 
 using maximum likelihood estimates (MLE)~\cite{Clauset:2009}.}
\label{tab:expo}
\begin{tabular}{|l|l|l|l|c|c|c|}
\hline

Dataset  & Description & Period & Events & Total human death & Range & $\nu$ \\ 
\hline

PRIO~\cite{PRIOdata}	& conflict  & 1946-2008 & 1957 & 4481398 
(lowest estimate) &  $> 1000$  & $1.54\pm0.06$  \\ \hline
CoW~\cite{CoWdata}	& war  & 1816-2007 & 662 & 40198386 & 
$> 3000$  & $1.63\pm0.03$  \\ \hline
UCDP~\cite{UCDPdata}	& battle  & 1989-2014 & 1306 & 1049906 & 
$> 1000$  & $1.64 \pm 0.07$  \\ \hline \hline

EMDAT earthquake~\cite{EMDATdata}	& earthquakes  & 1900-2013 & 
912 & 2564382 & $> 1000$  & $1.51\pm 0.05$  \\ \hline
EMDAT storm~\cite{EMDATdata}	& storms  & 1900-2013 & 
2653 & 1383677 & $> 100$  & $1.65\pm 0.03$  \\ \hline
EMDAT wildfire~\cite{EMDATdata}	& wildfires  & 1900-2013 & 
146 & 3665 & $> 50$  & $1.51\pm 0.01$  \\ \hline
EMDAT misc accidents~\cite{EMDATdata}	& miscellaneous accidents  & 
1900-2013 & 1062 & 62822 & $> 100$  & $1.81\pm0.12$  \\ \hline
EMDAT all disasters~\cite{EMDATdata}	& all disasters  & 1900-2013 
& 16295 & 38463886 & $> 5000$  & $1.48\pm0.03$  \\ \hline

\end{tabular}
\end{table}
%%%%%%%%%%%%%%%%%%%%%%%%%%%%%%%%%%%%%%%%%%%%%%%%%%%%

%%%%%%%%%%%%%%%%%%%%%%%%%%%%%%%%%%%%%%%%%%%%%%%%%%%%%%%%%%
%%%%%%%%%%%%%%%%%%%%%%%%%%%%%%%%%%%%%%%%%%%%%%%%%%%%%%%%%%%%

\section{Summary \& Discussions}
\label{sec:disc}
In summary, we find that the probability distributions $P(x)$ of event size 
$x$, measured by the number of human deaths in the corresponding event, goes as 
$P(x) \sim x^{-\nu}$ for the largest event sizes, with $\nu$ taking values 
between $1.5$ and $1.8$ for both man-made conflicts as well as natural 
disasters. In contrast, in the Guttenberg-Richter law of 
earthquakes~\cite{gutenberg1944frequency}, the probability of earthquakes 
releasing energy $E$ goes as $P(E) \sim E^{-\nu}$ with $\nu \simeq 
2$~\cite{christensen2002unified}.
Rainfall has also been reported to show a power law decay with $\nu \approx 
1.36$, measured in terms of volume of rain~\cite{peters2001complexity}.
Forest fires, as measured by the area of the land burned down, also shows 
similar characteristics, with power law decay exponent ranging between $1.3$ 
and $1.5$~\cite{malamud1998forest}.
Both the man-made conflicts and natural disasters, whether their size or 
magnitude is measured by released energy (in earthquakes), volume (rain), land 
area affected (forest fires) or by the fatal consequences given by the number 
of deaths, the size distribution has a power law tail $P(s) \sim s^{-\nu}$ with 
$\nu$ values roughly in the range $1.5-1.8$.
Table~\ref{tab:socexpo} summarizes the estimated values of $\nu$ for different 
classes of events.
The overall similarities suggest that 
physical modeling efforts may lend sufficient explanation for 
such statistical behavior of social event size distributions.

%%%%%%%%%%%%%%%%%%%%%%%%%%%%%%%%%%%%%%%%%%%%%%%%%%
\begin{table}[!htbp]
\caption{Various socio-economic events and their corresponding size exponent 
$\nu$ for the power law tail of the event size distribution.}
\label{tab:socexpo}
\begin{tabular}{|l|l|c|l|}
\hline

Event  & size measured in & $\nu$ & Reference  \\ 
\hline

City growth	& population count  & $1.8-2.2$  & 
Ref.~\cite{gabaix2004evolution}\\ \hline
Firm growth     & number of employees  & $\approx 2$ &  
Ref.~\cite{axtell2001zipf} \\ \hline
Income, wealth	& personal income, wealth  & $2-3$ & 
Ref.~\cite{chakrabarti2013econophysics} \\ \hline 
Firm bankruptcy & debt when firm is bankrupt & $\approx 2.1$ & 
Ref.~\cite{Fujiwara2004} \\ \hline
Earthquake	& energy released  & $\approx 2$ & 
Ref.~\cite{christensen2002unified} \\ \hline
Rainfall	& volume of water  & $1.36$ & 
Ref.~\cite{peters2001complexity}\\ \hline 
Forest fire     & land area burned & $1.3-1.5$ & Ref.~\cite{malamud1998forest} 
\\ \hline \hline

Man made events (wars \& conflicts) & death count  & $1.54-1.64$ & this paper, 
Fig.~\ref{fig:alldeaths} and Table.~\ref{tab:expo}  \\ \hline
Natural disasters & death count  & $1.48-1.81$ & this paper, 
Fig.~\ref{fig:alldeaths} and Table.~\ref{tab:expo} \\ \hline

\end{tabular}
\end{table}
%%%%%%%%%%%%%%%%%%%%%%%%%%%%%%%%%%%%%%%%%%%%%%%%%%%%

\begin{acknowledgments}
A.C. and B.K.C. acknowledge support from B.K.C.'s J. C. Bose Fellowship
Research Grant.
\end{acknowledgments}

\end{document}